\documentclass{IEEEtran}
\usepackage{cite}
\usepackage{amsmath,amssymb,amsfonts}
\usepackage{algorithmic}
\usepackage{graphicx}
\usepackage{textcomp}
\usepackage{xcolor}
\usepackage{booktabs,mathptmx,siunitx}
\usepackage{balance}
\usepackage[caption=false,font=normalsize,labelfont=sf,textfont=sf]{subfig}
\usepackage{multirow}

    \usepackage[compact]{titlesec}
    \titlespacing{\section}{0pt}{2ex}{1ex}
    \titlespacing{\subsection}{0pt}{1ex}{0ex}
    \titlespacing{\subsubsection}{0pt}{0.5ex}{0ex}
\def\BibTeX{{\rm B\kern-.05em{\sc i\kern-.025em b}\kern-.08em
    T\kern-.1667em\lower.7ex\hbox{E}\kern-.125emX}}
\begin{document}
\title{{\fontsize{24}{26}\selectfont{Communication\rule{29.9pc}{0.5pt}}}\break\fontsize{16}{18}\selectfont
On the Use of Impedance Detuning for Gastrointestinal Segment Tracking of~Ingestible Capsules}
\author{Erdem Cil, \IEEEmembership{Student Member, IEEE}, Icaro V. Soares, \IEEEmembership{Student Member, IEEE}, David Renaudeau, Ronan Lucas, Sema~Dumanli,~\IEEEmembership{Member, IEEE}, Ronan Sauleau,~\IEEEmembership{Fellow, IEEE}, and Denys Nikolayev, \IEEEmembership{Member, IEEE}
\thanks{Manuscript received July 13, 2022, revised November 25, 2022, accepted December 9, 2022. This work was supported 
in part by the French \textit{Agence Nationale de la Recherche} (ANR) under grant ANR-21-CE19-0045 (project ``MedWave''), in part by Région Bretagne (France) through the \textit{Stratégie d’attractivité durable} (SAD) project ``EM-NEURO'', in part by the BodyCAP Company, and in part by the European Regional Development Fund (ERDF) and Région Normandie (France). \textit{(Corresponding author: Denys Nikolayev.)}}
\thanks{E. Cil is with the Univ Rennes, CNRS, IETR~-- UMR 6164, FR-35000 Rennes, France and with BodyCAP, FR-14200 Hérouville St Clair, France.}
\thanks{I. V. Soares, R. Sauleau, and D. Nikolayev are with the Univ Rennes, CNRS, IETR~-- UMR 6164, FR-35000 Rennes, France (e-mail: denys.nikolayev@deniq.com).}
\thanks{D. Renaudeau is with the PEGASE, INRAE, Agrocampus-Ouest, 16 le clos, FR-35590 Saint-Gilles, France.}
\thanks{R. Lucas is with BodyCAP, FR-14200 Hérouville St Clair, France.}
\thanks{S. Dumanli is with the Department of Electrical and Electronics Engineering, Bogazici University, 34343 Istanbul, Turkey.}
}

\maketitle

\begin{abstract}
During their travel through the gastrointestinal tract, ingestible antennas encounter detuning in their impedance response due to varying electromagnetic properties of the surrounding tissues. This paper investigates the possibility of using this impedance detuning to detect in which segment of the gastrointestinal tract – stomach, small intestine, or large intestine – the capsule is located. Meandered dipole antennas operating in the 433 MHz Industrial, Scientific, and Medical Band are designed for this purpose. The antennas conform to the inner surface of 3D-printed polylactic-acid capsules with a shell thickness of 0.6 or 0.4 mm. The impedance response is first optimized numerically in a homogeneous cylindrical phantom with time-averaged electromagnetic properties. The magnitude and the phase of the reflection coefficient are then obtained in different tissues and compared with simulations and measurements. The experimental demonstration is carried out first using tissue-mimicking liquids and then in a recently deceased \textit{ex vivo} porcine model. The minimum change in the phase between different gastrointestinal tissues was determined to be around $10^\circ$ in the porcine model, indicating that the changes in the impedance response, particularly the changes in the phase, provide sufficient information to follow the position of the capsule in the gastrointestinal tract.
\end{abstract}

\begin{IEEEkeywords}
Conformal antennas, dipole antennas, impedance detuning, in-body, ingestible devices, \textit{in vivo} applications.
\end{IEEEkeywords}

\section{Introduction}
\label{sec:introduction}
 In recent years, ingestible devices have been widely used in diverse medical applications, such as monitoring of physiological data and wireless endoscopy \cite{katz2014implantable, 7855707, 6413167}. Due to the variety of possible applications and compact structure of these devices, the ingestible bioelectronics have the potential to substantially improve the diagnostics and therapies of the gastrointestinal (GI) tract \cite{9562451}.

Ingestible devices consist of several components integrated into the same capsule \cite{6304996}. The antenna is one of the key components as the quality of the communication link with external devices strongly depends on its performance \cite{nikolayevOptimalRadiationBodyimplanted2019, sipusInfluenceUncertaintyBody2021}. Therefore, different types of antennas, such as patch \cite{7902165, 8636757, iqbalScalpImplantableMIMOAntenna2021}, spiral \cite{5710972}, helical \cite{6930757, 8094934}, and differentially-fed antennas \cite{8288682, 9582771} have been implemented in ingestible capsules. Note that the ingestible antenna design is a challenging task due to the variety of loss mechanisms encountered in their operation. These losses can be classified as near-field, propagation, reflection, and detuning losses \cite{8606957, fangPathLossModels2021, nikolayevElectromagneticRadiationEfficiency2018, benaissaPropagationlossCharacterizationLivestock2021}. The latest corresponds to the possible changes in the impedance response of the antenna caused by the physical and electromagnetic (EM) variations in the surrounding biological environment as the capsule travels through the GI tract. Several studies in the literature examined these changes at different frequencies, such as 403.5 MHz \cite{6701534}, 433 MHz \cite{7481526, 8388259}, and 2.45 GHz \cite{bocanSimulatingModelingSensing2018, eucap2022}. Moreover, various robust ingestible antenna designs were proposed to mitigate this loss component \cite{9667256, 8607978, 8048539, 6547172, 9178280, 7934096, 8345685,nikolayevImmunetodetuningWirelessInbody2019}. 

Recent technological breakthroughs in terms of miniaturization, artificial intelligence, and power saving have enabled innovative companies to develop ingestible systems for monitoring physiological data~\cite{khaleghiRadioFrequencyBackscatter2019, bodycap}. The next generation of ingestible bioelectronic systems aims to provide information beyond this intent by performing accurate measurements all along the GI tract. For instance, the real-time localization of a capsule inside the GI tract can be a useful and innovative measurement to (i) optimize a device to investigate the middle of the track which is not accessible with probes (feeding tube, rectal probe), (ii) correlate a measurement and a position to determine the segment duration in order to identify specific pathologies such as gastroparesis, (iii) deliver efficiently an active substance at the right position. The smart capsules designed within this context can offer an ergonomic and accurate medical system. 

In this context, this paper investigates the possibility of using the impedance detuning caused by the varying EM properties of the GI tract to distinguish the GI tissues (i.e. stomach, small intestine, and large intestine) for localization purposes. For this investigation, impedance responses (both magnitude and phase of the reflection coefficient) of capsule-integrated meandered dipole antennas in different GI tissues are compared with simulations in numerical phantoms representing different GI tissues and with measurements in tissue-mimicking liquids as well as in an \textit{ex vivo} porcine model. To the authors' best knowledge, it is the first time in the literature where impedance responses of ingestible antennas are investigated for the purpose of GI segment tracking.

\begin{table}

\renewcommand{\arraystretch}{1.1}
\caption{EM Properties at 434 MHz and Concentrations (in grams) per 100 g for the Fabrication of the Tissue-Mimicking Liquids\label{tab:emproperties}}
\centering
\begin{tabular}{l S[table-format=2.2] S[table-format=2.2] S[table-format=2.2] S[table-format=2.2]}
\hline
&\textbf{Average} & \textbf{Stomach} & \textbf{Small Intestine} & \textbf{Large Intestine}\\
\hline
$\varepsilon_r$& 63 & 67.2 & 65.3 & 62\\
$\sigma$ (S/m) & 1.02 & 1.01& 1.92 & 0.87\\
Water& 53.96 & 59.13 & 54.59 & 52.97\\
Sugar & 44.6 & 39.7& 42.6 & 45.8\\
Salt & 1.44 & 1.17& 2.81 & 1.23\\
\hline
\end{tabular}
\vspace{-0.4cm}
\end{table}

\section{Numerical Study}
\subsection{Antenna Design and Modeling}
\begin{figure}
\centering
\includegraphics[trim={10cm 6.6cm 13.4cm 1.1cm},clip, width=\columnwidth]{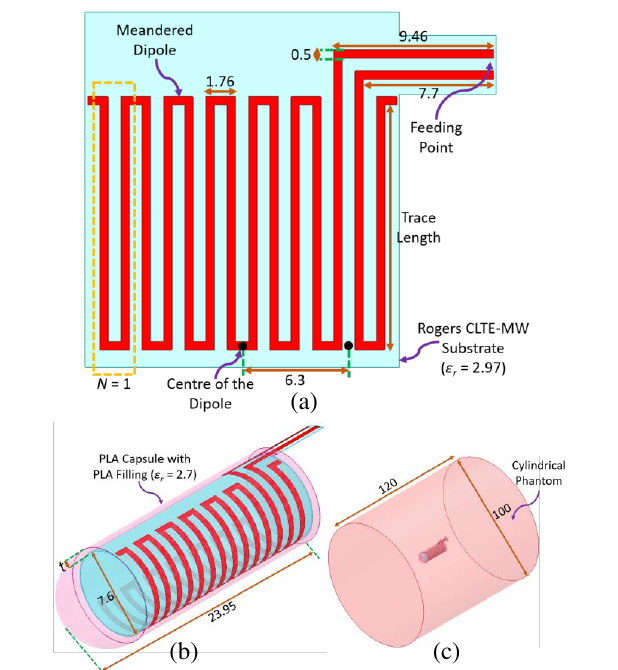}%
\caption{Model of the antenna, the capsule and the numerical setup (units: mm). (a) Meandered dipole antenna model. \textcolor{black}{(b)} Antenna conforming to the inner surface of the PLA capsule. (c) Capsule placed inside the cylindrical numerical phantom for simulations.}
\label{fig:models}
\end{figure}

A capsule-integrated antenna is designed for the ingestible applications. We choose a meandered dipole antenna for this study since its radiation mechanisms and the near-field behavior have been widely studied and better understood than more complex structures~\cite{8606957, 5991925}. The developed antenna [Fig. \ref{fig:models}(a)] operates in the 433 MHz Industrial, Scientific, and Medical (ISM) band. It is designed on a 100-\textmu m-thick Rogers CLTE-MW substrate ($\varepsilon_r$ = 2.97) \cite{rogers}, which conforms to the inner surface of the \textcolor{black}{biocompatible polylactic-acid capsule~\cite{DASILVA20189} (PLA, $\varepsilon_r = 2.7$, $\delta = 0.003$) as shown in Fig.~\ref{fig:models}(b)}. The capsule is  filled with a PLA cylinder to fix the antenna in its cylindrical shape. We studied the antenna detuning for two shells with  thickness of $t = 0.6$~mm or $t = 0.4$~mm. The capsule is placed in the middle of a cylindrical homogeneous phantom, as seen in Fig.~\ref{fig:models}(c). \textcolor{black}{The dimensions of the cylindrical phantom are depicted in Fig.~\ref{fig:models}(c). These dimensions ensure that the near-field of the antennas is confined within the phantom, i.e., the impedance responses of the antennas remain stable when the dimensions of the phantom are further increased.} The trace length is numerically optimized for the two shell thicknesses in the phantom having time-averaged EM properties of the GI tract ($\varepsilon_r = 63$, $\sigma = 1.02$~S/m). The detailed description of the time-averaged phantom can be found in \cite{eucap2022, no2}. The optimized values for the trace length are 16.9~mm for $t = 0.6$~mm and 14.5 mm for $t = 0.4$~mm. The values of the other parameters are the same for two shell thicknesses and are shown in Fig. \ref{fig:models}(a). The parameter $N$ indicates the number of turns as explained in \cite{6130586} and is equal to 7 for both models. \textcolor{black}{Note that, as shown in Fig. \ref{fig:models}(a), an offset feed (6.3 mm from the center) is used to feed the dipoles, making one dipole arm longer. Although a free-space center-fed dipole can be easily matched to 50~$\Omega$, in-body capsule application requires miniaturizing the antenna and taking account loading by tissues. Offsetting the feed is one of the techniques to match the in-body dipole antenna to 50~$\Omega$~\cite{4812248}}.

\subsection{Numerical Results}
\begin{figure}
\centering
\includegraphics[scale = 1.06]{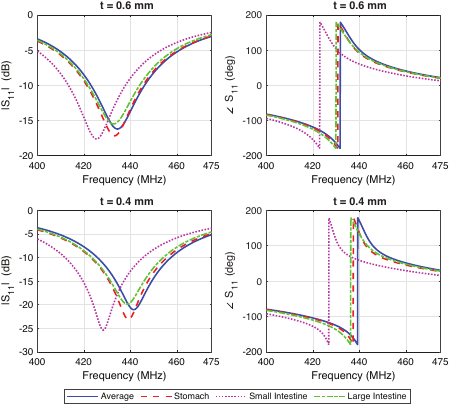}%
\caption{Simulation results for the magnitude and the phase of the reflection coefficient obtained in numerical phantoms representing different GI tissues for two shell thicknesses.}
\label{fig:simresults}
\vspace{-0.7cm}
\end{figure}

After optimizing the antennas, the possibility of using the impedance detuning is examined by simulating the antennas in numerical phantoms representing 3 tissues in the GI tract: stomach, small intestine, and large intestine. The EM properties of these tissues at 434 MHz are tabulated in Table \ref{tab:emproperties} \cite{gabriel}. The simulated magnitude and phase of the reflection coefficient are shown in Fig. \ref{fig:simresults} \textcolor{black}{and the values at 434 MHz are tabulated in Table \ref{tab:numbers}.} From the results, it can be seen that the minimum difference in the magnitude observed between 3 GI tissues at 434 MHz is 0.9~dB for $t = 0.6$~mm and 1.1 dB for $t = 0.4$~mm. As for the phase, it is $16.9^\circ$ for $t = 0.6$~mm and $19.7^\circ$ for $t = 0.4$~mm. As these values are sufficiently large to distinguish the tissues, it can be stated that the change in the impedance response of the antenna in different GI tissues can be utilized to determine the segment in which the capsule is located. Note that it is more convenient to track the changes in the phase for the intended purpose, as they are larger and more consistent than the changes in the magnitude. \textcolor{black}{The phase values are predominantly affected by the changes in the conductivity of surrounding tissues. The effect of the relative permittivity on the phase is insignificant. In the conductivity range of 0.8--2 S/m, which covers all GI tissues at 434~MHz, the conductivity is inversely proportional to the phase angle. For the 0.4~mm shell, the phase follows the law $\phi = -131.7\sigma + 309$ ($R^2 = 0.932$), and for the 0.6~mm, $\phi = -75.1\sigma + 217$ ($R^2 = 0.948$).}

\begin{table}
\vspace{-0.4cm}
\setlength{\tabcolsep}{2.5pt}
\renewcommand{\arraystretch}{1.1}
\caption{\textcolor{black}{Simulated and Measured (Phantoms) Reflection Coefficients\label{tab:numbers} (ST: Stomach, SI: Small Intestine, LI: Large Intestine)}}
\centering
\begin{tabular}{l l S[table-format=-3.1] S[table-format=-2.1] S[table-format=-3.1]|  S[table-format=-3.1] S[table-format=-2.1] S[table-format=-3.1] }
\hline
&&\multicolumn{3}{c|}{\textbf{t = 0.6~mm}}&\multicolumn{3}{c}{\textbf{t = 0.4~mm}}\\
&&\textbf{ST} & {\textbf{SI}} & \textbf{LI}&\textbf{ST} & {\textbf{SI}} & \textbf{LI}  \\
\hline
\multirow{2}{*}{Sim.}&$|S_{11}|$ (dB) &-15.3  &-12.3 &-16.2 &-17.1 &-18.2 &-15.2 \\
& $\angle{S_{11}}$ (deg.) &139.8 &80.6 &156.7&-160 &78.4 &-140.3   \\
\hline
\multirow{2}{*}{Phant.}& $|S_{11}|$ (dB)&-21.7 &-21.9 &-21.3&-20.2&-19.9 &-21.4   \\
&$\angle{S_{11}}$ (deg.) &93.6 & 98.9 &55.2&155.5&131&171.8 \\
\hline
\end{tabular}
\vspace{-0.4cm}
\end{table}

\section{Prototyping and Measurements}
The measurements were carried out using two different measurement setups. In the first setup, tissue-mimicking liquids were used. In the second setup, \textit{ex vivo} measurements were performed using a porcine model.
\subsection{Prototyping}
\begin{figure}
\centering
\includegraphics[width=0.9\columnwidth]{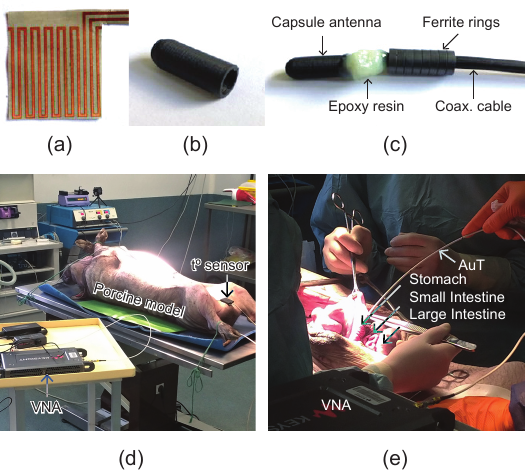}%
\caption{Different parts of the prototype and the porcine model. (a) Printed antenna. (b) 3D-printed PLA capsule. (c) Final antenna under test (AuT) with the ferrite rings. (d) General view of the porcine model used in the \textit{ex vivo} measurements. (e) AuT placed inside the porcine model. }
\label{fig:prototyping}
\vspace{-0.5cm}
\end{figure}

The antennas and 3D-printed capsules were fabricated for the measurements, as shown in Fig. \ref{fig:prototyping}(a) and Fig. \ref{fig:prototyping}(b), respectively. The antennas were soldered to the coaxial cable and placed inside the capsules. The side of the capsules was covered with Araldite 2012 epoxy resin to make a watertight seal. As mentioned previously, the main purpose of this study is to compare the impedance responses of the antennas in different GI tissues. However, with the introduction of the coaxial cable, the EM changes in the environment around the cable also affect the impedance responses, decreasing the accuracy of the comparison. Therefore, ferrite rings, which help eliminate this undesired effect by attenuating the magnetic fields created between the cable and the antenna, were placed at the antenna end of the cable. Note that the cable affects the operation of the antennas even with the use of ferrite rings. However, the ferrite rings stabilize this effect in different tissues by making it independent of the surrounding environment, which is sufficient for the aim of this work. The complete prototype with the ferrite rings is shown in Fig. \ref{fig:prototyping}(c).  
\vspace{-0.4cm}
\subsection{Measurements in Tissue-Mimicking Liquids}
\begin{figure}

\centering
\includegraphics[scale = 1.06]{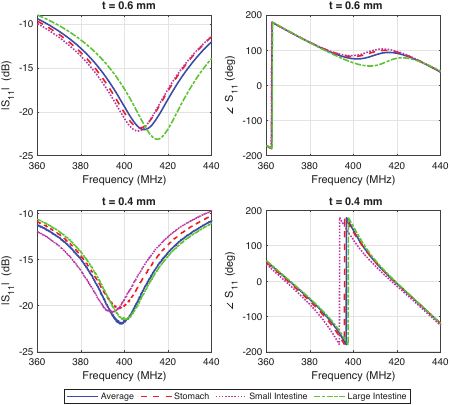}%
\caption{Measurement results for the magnitude and the phase of the reflection coefficient obtained in tissue-mimicking liquids for two shell thicknesses.}
\label{fig:phantomresults}
\vspace{-0.7cm}
\end{figure}

For the first measurement setup, tissue-mimicking liquids were fabricated with deionized water, sugar, and salt using the method proposed in \cite{8048539}. The amount of each ingredient used for 100~g tissues is tabulated in Table \ref{tab:emproperties}. The EM properties of the fabricated liquids were validated at 434~MHz with a SPEAG DAK-12 probe \cite{probe}.
Fig. \ref{fig:phantomresults} shows the measurement results in the prepared liquids. It can be observed that the operating frequency in the liquid with time-averaged EM properties is shifted to 408.6 MHz for $t = 0.6$~mm and 398.3~MHz for $t = 0.4$~mm due to the effect of the cable. \textcolor{black}{The magnitude and the phase of the reflection coefficient at these resonant frequencies are tabulated in Table \ref{tab:numbers}.} As can be seen, the minimum difference in the phase observed between 3 GI tissues is $5.3^\circ$ for $t = 0.6$~mm and $16.3^\circ$ for $t = 0.4$~mm. Similar to the simulation results, these values are sufficiently large to identify the tissue in which the capsule is located, supporting the idea presented in this work.
\subsection{Ex vivo Measurements}
Experimental validation of the proposed EM-sensing with ingestible capsules requires measurements using recently deceased tissues as the EM properties might quickly evolve post-mortem \cite{gabriel2}. Thus, \textit{ex vivo} measurements on a porcine model were performed at the experimental facilities of the Unité Expérimentale Physiologie et Phénotypage des Porcs in Saint-Gilles, France \cite{inrae} [Fig. \ref{fig:prototyping}(d)].

After premedication with ketamine (15 mg/kg IM), a 50-kg live body-weight Yucatan miniature pig was euthanized by intravenous injection of 5 ml of T61 – a nonbarbiturate, non-narcotic combination consisting of 3 compounds: embutramide (anesthetic), mebenzonium iodide (paralytic), and tetracaine hydrochloride (potent local anesthetic). Once euthanized, the animal was placed in dorsal recumbency. An incision (about 30 cm) was made in the lower abdominal wall ventral median following the linea alba starting 100 mm behind the breastbone, opening the abdominal cavity. The stomach (M1), the apical part of the duodenum (M2), the terminal part of the ileum (at the beginning: M3 and at the end: M4), and the terminal part of the colon (M5) were located and exteriorized. At each location, 1 cm incisions were made and the prototypes were inserted in 10 cm depth as shown in Fig. \ref{fig:prototyping}(e).

\begin{figure}
\centering
\includegraphics[scale = 1]{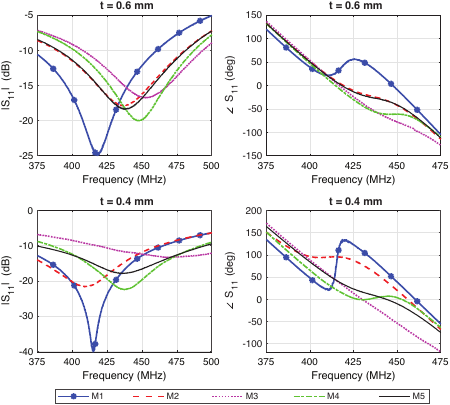}%
\caption{Measurement results for the magnitude and the phase of the reflection coefficient obtained with the porcine model for two shell thicknesses.}
\label{fig:realtissueresults1}
\vspace{-0.7cm}
\end{figure}

\begin{figure}
\centering
\includegraphics[scale = 1]{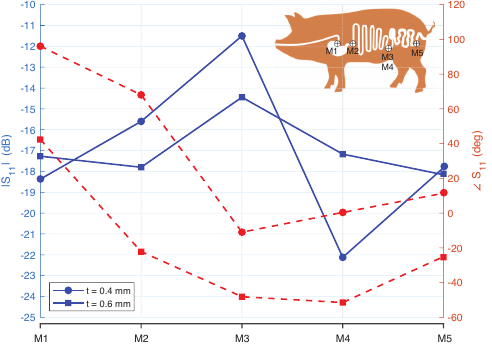}%
\caption{The values of the magnitude and the phase of the reflection coefficient at 434 MHz measured using the porcine model. The straight lines show the magnitude and the dashed lines show the phase.}
\label{fig:realtissueresults2}
\vspace{-0.7cm}
\end{figure}

 Fig. \ref{fig:realtissueresults1} shows the measurement results obtained using the porcine model. As can be seen, the change in the magnitude and phase values in different tissues is greater compared to the simulations and measurements in the liquids. The responses in different tissues become more distinguishable with a more realistic setup, as the real tissues are heterogeneous and contain air gaps, which contributes to the differences in the responses. Furthermore, Fig. \ref{fig:realtissueresults2} \textcolor{black}{and Table \ref{tab:animalnumbers}} illustrate the magnitude and phase values measured at 434 MHz with the porcine model. It can be observed that the values change significantly depending on the tissue. This indicates that it is possible to distinguish the tissues by tracking the impedance response. \textcolor{black}{Moreover, the graphs obtained with two different antennas have similar tendencies. For instance, for both antennas, the phase is decreasing from M1 to M2, and it is increasing from M4 to M5. These results show the consistency of the measurements and the stability of the changes, supporting the hypothesis presented in this study.}

\begin{table}
\setlength{\tabcolsep}{6pt}
\renewcommand{\arraystretch}{1.1}
\caption{\textcolor{black}{Reflection Coefficient Obtained with the Porcine Model}\label{tab:animalnumbers} }
\centering
\begin{tabular}{ l S[table-format=-3.1] S[table-format=-3.1]| S[table-format=-3.1] S[table-format=-3.1] }
\hline
&\multicolumn{2}{c|}{\textbf{t = 0.6~mm}}&\multicolumn{2}{c}{\textbf{t = 0.4~mm}}\\
& {$|S_{11}|$ (dB)} & {$\angle{S_{11}}$ (deg.)}&{$|S_{11}|$ (dB)} & {$\angle{S_{11}}$ (deg.)} \\
\hline
{\textbf{M1}} &-17.3  &42.4 &-18.4 &96.1 \\
{\textbf{M2}} &-17.8  &-22.2 &-15.6 &68.1 \\
{\textbf{M3}} &-14.4  &-48.1 &-11.5 &-11 \\
{\textbf{M4}} &-17.2  &-51.5 &-22.1 &0.4 \\
{\textbf{M5}} &-18.2  &-25 &-17.8 &11.4 \\
\hline
\end{tabular}
\vspace{-0.7cm}
\end{table}

\section{Conclusion and Perspectives}
This paper studied the feasibility of using the impedance detuning of ingestible antennas to track the segment-level location of the GI tract. For this, we compared the impedance responses of ingestible antennas in different GI tissues numerically and experimentally in tissue-mimicking liquids and porcine model. The results show that the location of the capsule can be followed by tracking the changes in the impedance response, in phase in particular. 

\textcolor{black}{This study validates the proposed methodology and leads to the development and pre-clinical validation \textit{in vivo} of a fully-wireless model. In practice, the phase shift measurement at a single frequency is straightforward and can be implemented using off-the-shelf miniature and low-power RF components~\cite{nikolayevBiotelemetryDeviceThat2021a}. The averaged relative shift in the phase -- a derivative of the surrounding tissue EM properties -- can be quantified \textit{a posteriori} as a function of the tissue where the capsule is located and used for the tracking purposes. Note that the phase shift between different tissues can be adjusted by increasing the near-field coupling of the antenna with the tissues. Optimal design of such multiplexed sensor--antenna therefore requires a compromise between the sensing and radiation performance. This is an open research question that requires further investigation and defines our future work.} Another perspective direction is to implement biodegradable resonators (for instance, Mg) on the outside part of the capsule as an additional indicator. Such a resonator will rapidly degrade in the stomach due to its acidic nature and hence can be used to distinguish the stomach from the other segments. The real-time localization of capsules inside the GI tract can be useful to improve diagnostics as well as to deliver drugs and therapies more efficiently. 

\section*{Acknowledgement}
The authors would like to thank Sébastien Moussay and Estelle Blond from BodyCAP, without whom this research would have been impossible, Christophe Guitton for the manufacturing of prototypes, Frédéric Boutet and Laurent Le Coq for helping with the antenna measurements, and Céline Gantier for the assistance with the \textit{ex vivo} demonstration.
\balance
\bibliographystyle{IEEEtran}
\bibliography{References}

\end{document}